\documentclass[conference]{IEEEtran}
\IEEEoverridecommandlockouts
\usepackage{dirtytalk}
\usepackage{multirow}  
\usepackage{multicol}  
\usepackage{makecell}
\usepackage{subfigure}
\usepackage{hyperref}

\usepackage{listings}
\usepackage{fancyvrb}
\usepackage{framed}
\usepackage[listings,skins]{tcolorbox}
\usepackage[skipbelow=\topskip,skipabove=\topskip]{mdframed}
\mdfsetup{roundcorner=0}
\usepackage{amsmath,amssymb,amsfonts}
\usepackage{algorithmic}
\usepackage{graphicx}
\usepackage{textcomp}
\usepackage{xcolor}
\usepackage[hang,flushmargin]{footmisc}
\def\BibTeX{{\rm B\kern-.05em{\sc i\kern-.025em b}\kern-.08em
    T\kern-.1667em\lower.7ex\hbox{E}\kern-.125emX}}
\begin{document}

\title{Improving Quality of a Post's Set of Answers in Stack Overflow}
\author{
\IEEEauthorblockN{Mohammadreza Tavakoli}
\IEEEauthorblockA{\textit{Computer Engineering} \\
\textit{Sharif University of Technology}\\
rtavakoli@ce.sharif.edu} \and
\IEEEauthorblockN{Maliheh Izadi}
\IEEEauthorblockA{\textit{Computer Engineering} \\
\textit{Sharif University of Technology}\\
maliheh.izadi@sharif.edu} \and
\IEEEauthorblockN{Abbas Heydarnoori}
\IEEEauthorblockA{\textit{Computer Engineering} \\
\textit{Sharif University of Technology}\\
heydarnoori@sharif.edu}
}

\maketitle

\begin{abstract}
Community Question Answering platforms such as Stack Overflow help a wide range of users solve their challenges on-line. As the popularity of these communities has grown over the years, both the number of members and posts have escalated.
Also, due to the diverse backgrounds, skills, expertise, and viewpoints of users, each question may obtain more than one answers.
Therefore, the focus has changed toward producing posts that have a set of answers more valuable for the community as a whole, not just one accepted-answer aimed at satisfying only the question-asker.
Same as every universal community, a large number of low-quality posts on Stack Overflow require improvement.
We call these posts “\textit{deficient}”, and define them as posts with questions that either have no answer yet or can be improved by other ones.
In this paper, we propose an approach to automate the identification process of such posts and boost their \textit{set of answers}, utilizing the help of related experts.
With the help of 60 participants, we trained a classification model to identify deficient posts by investigating the relationship between characteristics of 3075 questions posted on Stack Overflow and their need for better answers set.
Then, we developed an Eclipse plugin named SOPI and integrated the prediction model in the plugin to link these deficient posts to related developers (in terms of their development context and expertise area) and help them improve the answer set.
We evaluated both the functionality of our plugin and the impact of answers submitted to Stack Overflow with the help of 10 and 15 expert industrial developers, respectively.
Our results indicate that decision trees, specifically the J48 algorithm, predicts a deficient question better than the other methods with 94.5\% precision and 90.3\% recall.
We conclude that not only our plugin helps programmers contribute more easily to Stack Overflow, but also it improves the quality of existing answers.
\end{abstract}

\section{Introduction}\label{sec:intro}
Stack Overflow (SO) is a successful example of Community Question and Answering (CQA) platform, specifically designed for solving programmers' and software engineers' challenges.
According to the 2019 global survey of SO, approximately 50M people visit SO monthly. Professional developers and university-level students cover 42\% of these visits\footnote{SO Survey 2019: \url{https://insights.stackoverflow.com/survey/2019}}.
As of October 2019, about 18M questions have been posted on SO. Approximately, users have posted 28M answers to these questions, from which only 9.6M answers have been accepted.
SO also reports users post on average 7.9K new questions and 8.1K new answers each working day\footnote{This data has been updated on Oct, 14th, 2019 using Stack Exchange API}.
Although some questions have no answer, others receive numerous ones. This shows the difference between users' views and usage types, thus emphasising the importance of constructing high-quality answer sets.
 
On the other hand, due to the fast-growing community of SO and the inability to carefully assess the content being published, a large number of poorly edited answers, incorrect or obsolete ones, incomplete code snippets, too-specific or too-general answers, misleading information, and other perils exist among the answers set of posts.
Posts on SO, as other software artifacts, need to change and evolve over time. 
Edits include fixing code snippets' bugs, updating them to work with a more recent version of a library, or addressing the issue from other perspectives \cite{zhang2019empirical,baltes2018evolution}.
The numbers confirm this notion as 13.9M SO posts have been edited after their creation time, 19K of them more than ten times \cite{baltes2018evolution}.
To illustrate a case of such need for attentive maintenance and evolution overtime, we use a SO post as an example\footnote{\url{https://stackoverflow.com/questions/1302605}}.
This post contains a question that potentially can be improved with new answers.
``How do I convert from Int to Long in Java?"
This post has been viewed 581K times and has 13 answers. From this answers set, our SOPI generated answer has obtained 11 up-votes so far. 
Although this question has been asked 10 years ago, it is still active, which indicates the continuous interest in the topic and the need for better answers.
However, programmers are not always motivated to answer or update CQA websites inherently \cite{treude2011programmers}, since it is a time-consuming activity and needs focus.
So we can motivate them either through rewarding mechanisms, or by simplifying the answering process as much as possible \cite{liu2017identifying}.
Although SO has used various motivating approaches such as awarding badges, studies have shown these approaches have not been successful enough in the task of improving overall quality of SO's posts \cite{wang2018users}.
Furthermore, Treude and Robillard \cite{treude2017understanding} studied the quality of code examples on SO and reported less than half of them have self-explanatory code fragments and therefore require improvement.
Considering that existing code snippets may be faulty, hard to understand, or have a low quality which is misleading in new user-queries \cite{ford2018we}, it is extremely important to correct errors and improve the quality of code snippets.

Therefore, SO needs a more efficient and easier solution for continuous improvement and control over the content.
We believe this is possible through exploiting the community itself to fix the existing flaws, improve answers, update obsolete answers, generalize too-specific answers, address issues from different views, enrich available code snippets, etc. collectively.
Different studies have addressed this problem from various aspects and tried to propose solutions for identifying and improving the quality of SO posts \cite{tavakoli2016improving, ponzanelli2014prompter}.
To the best of our knowledge, while previous work has focused on improving the process of finding better answers or speeding up the process, none have tried to identify questions that need better answers and enhance the quality of such posts.
We set the criteria for posts that need improvement as posts that have no answer or have some sort of an answer, but can have better or extended ones in terms of quality, level of expertise, details, etc. to satisfy different programmer's needs.
For instance, while an answer can solve an expert's question, it can be very hard for a novice to comprehend. 
Contrary to that, some posts have highly-detailed answers, whereas a group of the audience may prefer a brief answer to solve their problems faster. 
We call these posts \emph{``deficient"}.
Our contributions include:
\begin{itemize}
\item Addressing a practical problem, i.e., improving the quality of a post's answers set using developers' coding context.
\item Training a classification model to automatically identify deficient posts of SO.
\item Developing a tool, SOPI, to facilitate and accelerate the process of answering and contributing to SO with minimal distraction.
\item Three-fold evaluation involving 85 industrial Java developers.
\end{itemize}
Our study consists of four steps.
In step 1, we performed an exploratory analysis on 3075 questions from SO with the help of 60 developers to understand the properties of deficient posts. For their For the assessment, developers considered four measures of answers including \textit{completeness}, \textit{correctness}, \textit{conciseness} and \textit{comprehensibility}; then we investigated the relationship between 11 properties, such as their \textit{score}, \textit{has accepted answers}, and \textit{view count} of posts with their need for improvement.
From  these  properties,  we  chose  the  most  influential  features  and  trained  three  classifiers  to predict whether a post needs enhancement.
We used Neural Network (NN), Decision Tree (DT), and Support Vector Machine (SVM) models to classify the posts.
According to our evaluation results, J48, a DT classifier, best predicted these posts with 94.5\% precision and 90.3\% recall.
For step 2, 3, and 4 we developed an Eclipse plugin named \emph{SOPI} (Stack Overflow Post Improver).
SOPI first finds related questions based on a developer's programming context, then it filters them based on the developer's area of expertise using the history of selected user in SO as the second stage of filtration. 
SOPI then predicts which questions are deficient as the last filtration and based on the preference of the developer (suggestions frequency), it prompts the developer to improve the content of selected posts.
To make it even more simple and time-efficient, SOPI recommends a code snippet according to the programmers' code and the question's code snippets \cite{tavakoli2016improving}.
Finally, the new confirmed answer is submitted to SO with minimum interruption.
Therefore, our research questions include:
\begin{itemize}
\item RQ1: Can we train a model to predict deficient posts?
\item RQ2: Does SOPI facilitate the process of answering SO questions?
\item RQ3: Do the posted answers improve the quality of posts?
\end{itemize}

\section{Properties of Deficient Posts}\label{sec:prop}
Here we describe the process of identifying a set of properties for deciding whether a post needs improvement or not.
\subsection{Data Collection}
Using Stack Exchange API\footnote{\url{https://api.stackexchange.com/}}, we randomly collected data of 3600 posts with the last activity in the time range of 2013 to 2019.
These questions have a ``Java" tag and are not deleted or closed in SO. We also retrieved 11 properties that are available for collection using the SO API to investigate the necessity of enhancement for a post.
These properties are:
\begin{itemize}
\item \emph{Has Accepted Answer (HAA):} Whether the post has an answer that satisfies the asker?
\item \emph{Answer Count (AC):} The number of answers in a post.
\item \emph{Score (S):} The score of a question based on up-votes and down-votes.
\item \emph{Sum of the Answer Scores (SAS):} The sum of set of answers' scores based on their up-votes and down-votes.
\item \emph{View Count (VC):} The view count of a post.
\item \emph{The Ratio of Sum of Scores to the View Count (SSVC):} The sum of answer scores divided by the view count of a post.
\item \emph{Comment Count (CC):} The number of comments on a question.
\item \emph{Favorite Count (FC):} The number of users who liked the question.
\item \emph{Average of Comment Count (ACC):} The average number of comments on the set of answers.
\item \emph{Average of the Answerer Reputation (AAR):} The average reputation of answerers.
\item \emph{Asker Reputation (AR):} The reputation of the asker.
\end{itemize}
Then, we assigned these posts to 60 experts with a degree in Computer Science and three months to five years of industrial experience. Furthermore, each post was covered by at least three participants. That is, each participant labeled 180 posts.
We asked them to check whether these posts require more appropriate answers, considering the following four criteria which are utilized in previous studies as well \cite{tavakoli2016improving}:
\begin{itemize}
\item \emph{Completeness:} Whether the post has at least one answer complete enough to solve the problem.
\item \emph{Conciseness:} Whether the post has at least one concise answer without non-informative parts.
\item \emph{Correctness:} Whether all answers to the question are without errors.
\item \emph{Comprehensibility:} Whether the answers are easy to understand.
\end{itemize}
We first held a three-hour meeting with the participants and established basic definitions to reach a common ground.
Then, we asked the participants to label a question with the tag ``YES" if at least one of the above conditions is not satisfied (regarding the four criteria).
Otherwise, if they think that the post's quality measures are fulfilled concerning all four aspects, they should label it as ``NO".
That is if a post needs further improvement, it is considered deficient in its current state, and subsequently is labeled ``YES". Otherwise, it is labeled ``NO".
To prevent biasing the participants, posts' information such as their score, has accepted answer, score, etc. is hidden during the manual labeling. 
We labeled a post \emph{YES}, if it had three \emph{YES} tags which means it is deficient and needs improvement.
Similarly, if a question had three ``NO" tags, we set its label to \emph{NO}.
It is worth mentioning we omitted 525 posts after the first round of labeling because we could not collect three labels for them. In the end, 3075 posts remained.
For other cases (when a post had two ``YES" tags and one ``NO" tag or had one ``YES" tag and two ``NO" tags), we created a five-member group and they discussed each post together.
After the negotiation, all the posts obtained at least four similar labels, so we labeled them based on the majority of the votes\footnote{The labeled data is available at \url{https://github.com/MalihehIzadi/SOPI_stackoverflow_answer_quality}}.

Below, we provide some information about the sample.
\begin{itemize}
\item About 66\% of posts have an accepted answer.
\item A third of posts have at least two answers. 
\item Half of posts have at least one score.
\item SAS for about 70\% of posts is between 1 to 8.
\item About 75\% of posts are viewed between 18 to 2000 times. Also, the maximum view of the sample count is 159,745.
\item The maximum SSVC is around 25\%. Also, for around 75\% of the posts, the ratio is less than 1\%.
Note that some posts have negative values for SSVC since sum of their scores is negative.
\item About 33\% of posts do not have any comments. However, 60\% of them have between one to seven comments.
\item About 66\% of posts do not have any FC.
\item About 57\% of posts on average have one to 11 comments on their answers.
\item The AAR of 50\% of posts is between 12K to 32K.
\item The AR in only 25\% of posts is more than 1K.
\end{itemize}
\subsection{Comparing the Properties of the Two Groups}
\begin{figure*}
    \centering
    \subfigure [Has Accepted Answer (HAA)]
    {\fbox{ \includegraphics[width=2.1in]{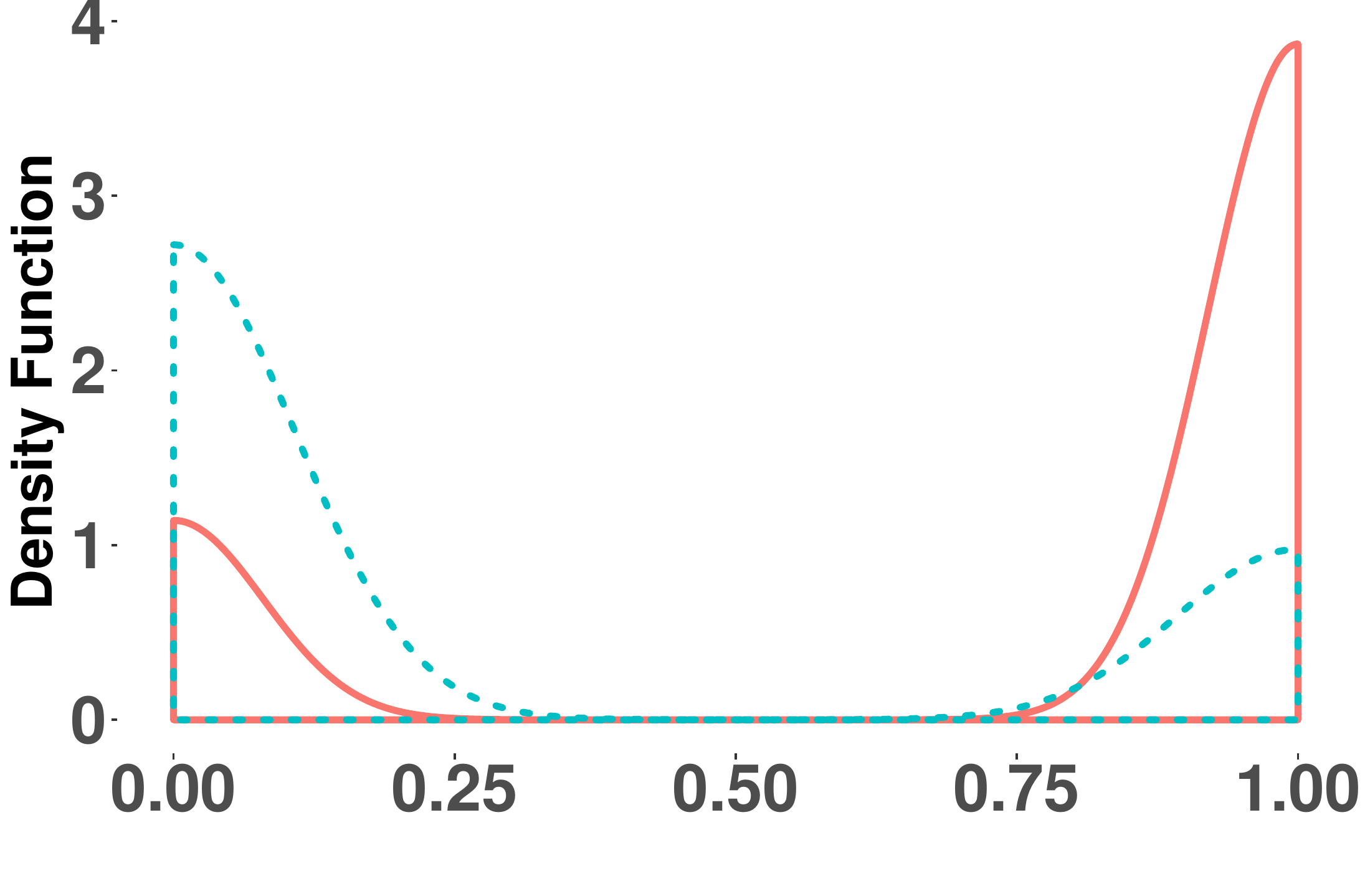}}
        \label{fig:his_has} }
    \subfigure[Answer Count (AC)]
    {\fbox{ \includegraphics[width=2.1in]{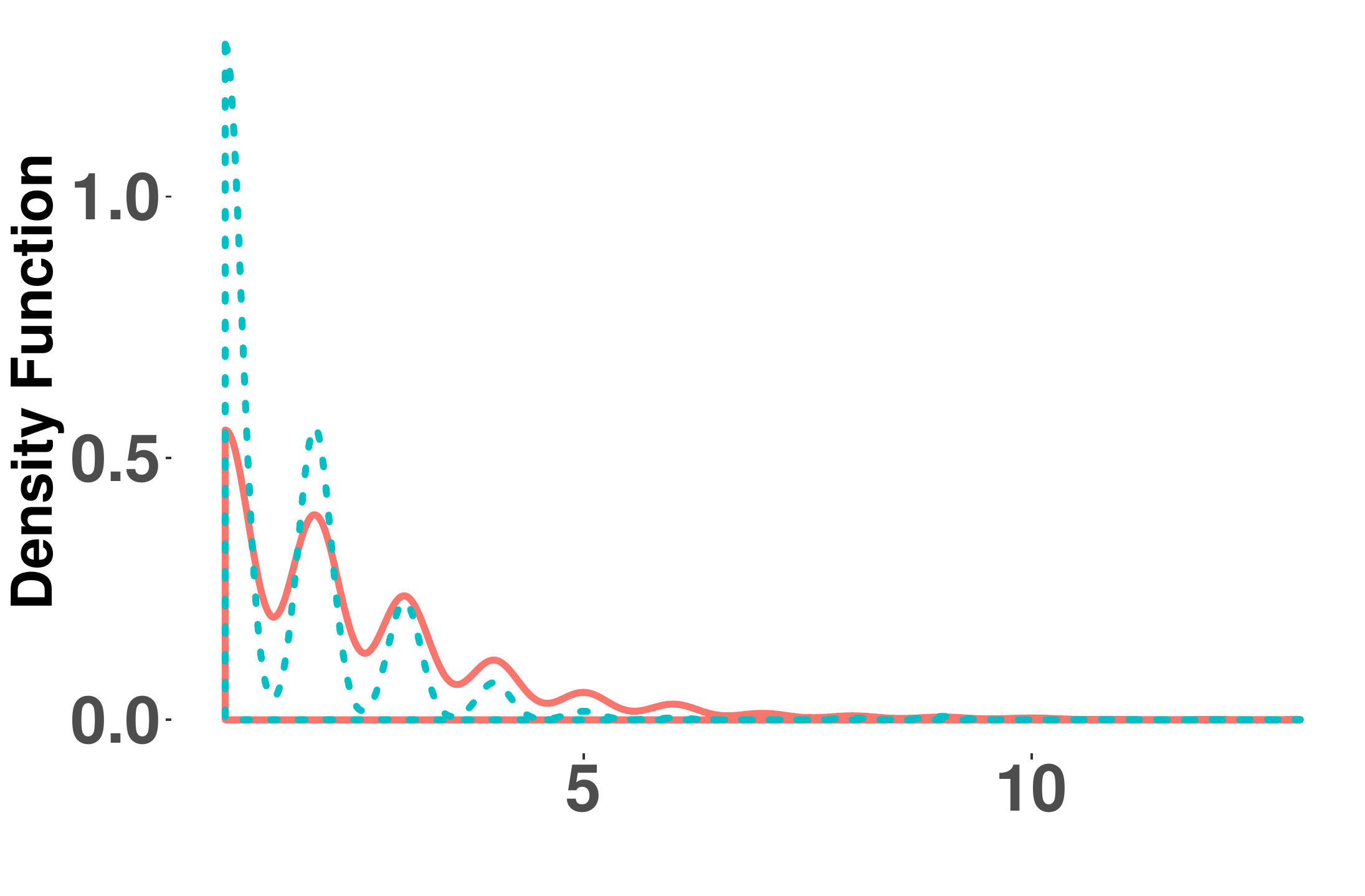}}
        \label{fig:his_ansCou}}
    \subfigure[Score (S)]
    {\fbox{\includegraphics[width=2.1in]{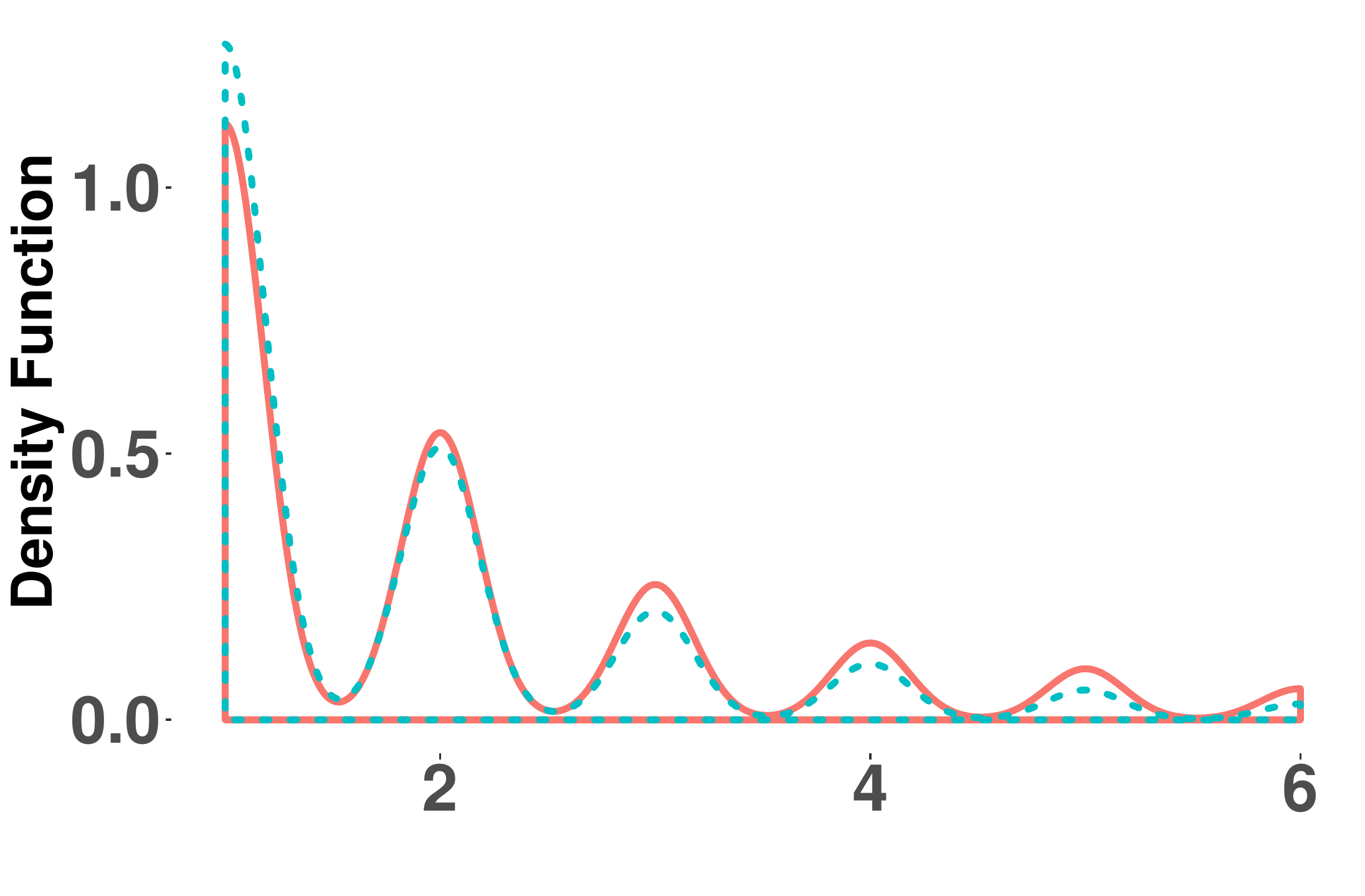}}
        \label{fig:his_sco}}
        
     \subfigure[Sum of the Answers' Score (SAS)]
        {\fbox{\includegraphics[width=2.1in]{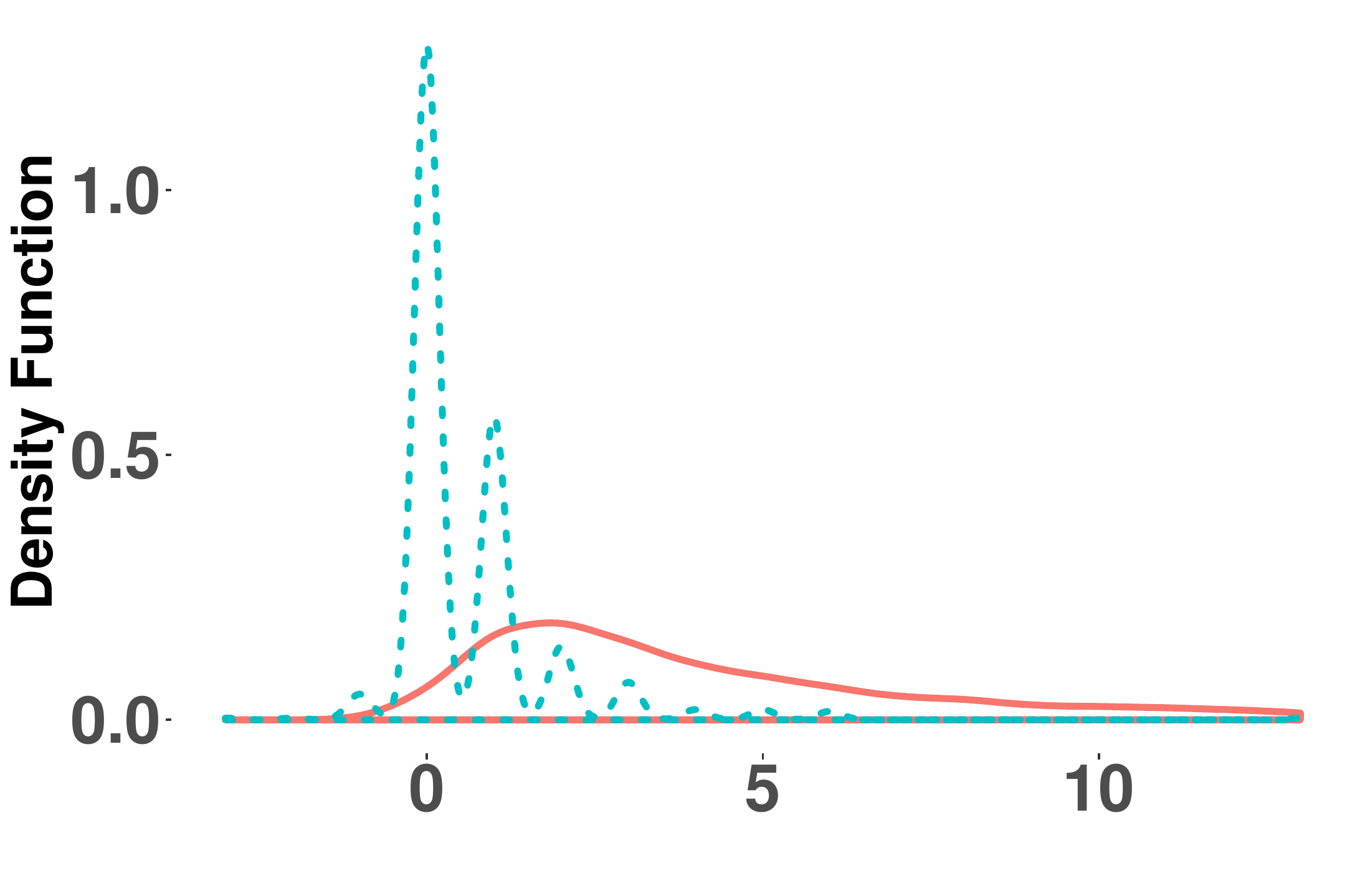}}
            \label{fig:his_sumSco}}
    \subfigure[View Count (VC)]
    {\fbox{\includegraphics[width=2.1in]{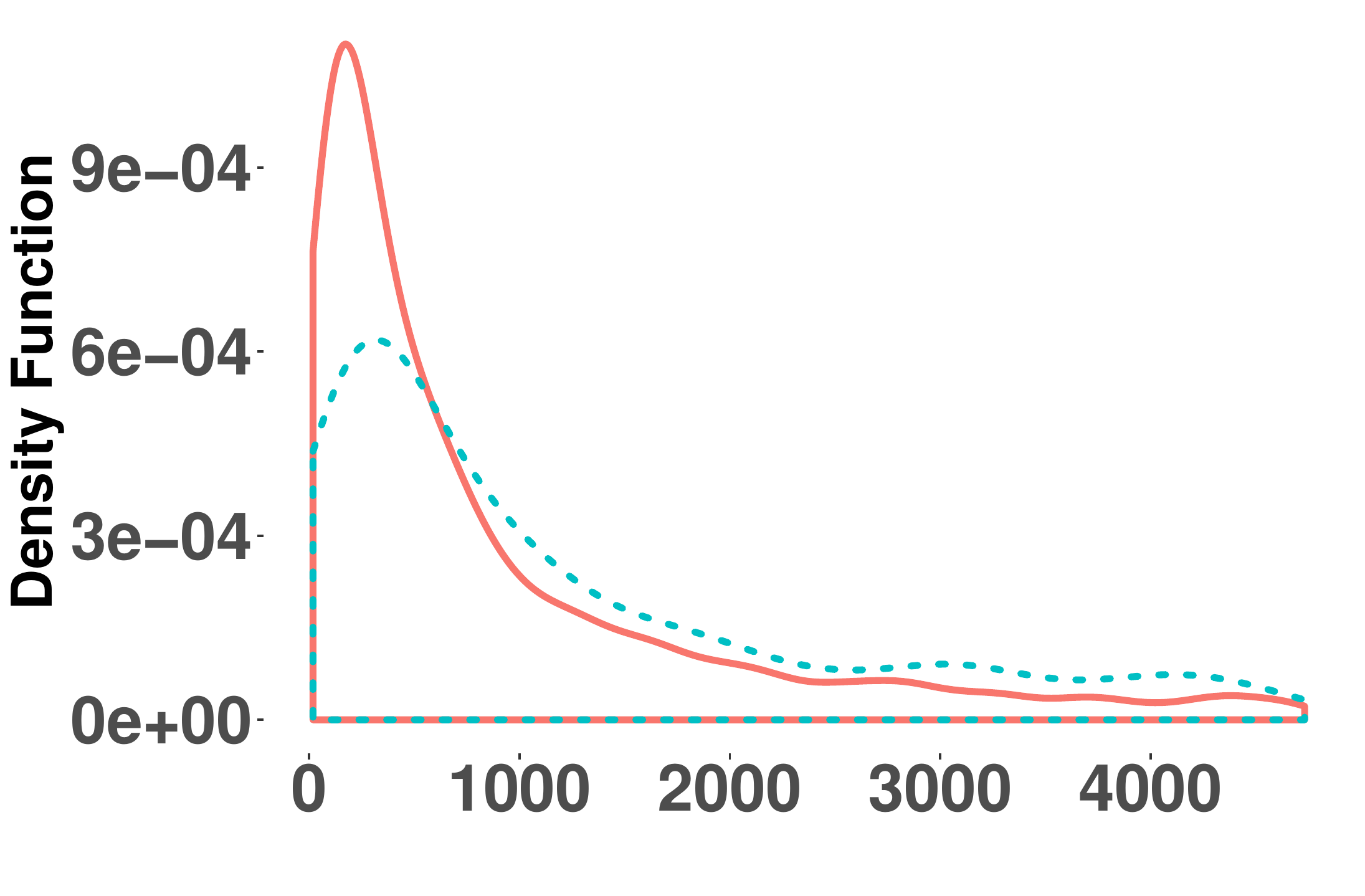}}
        \label{fig:his_viewCou}}
     \subfigure[Sum of the Scores/View Count (SSVC)]
        {\fbox{\includegraphics[width=2.1in]{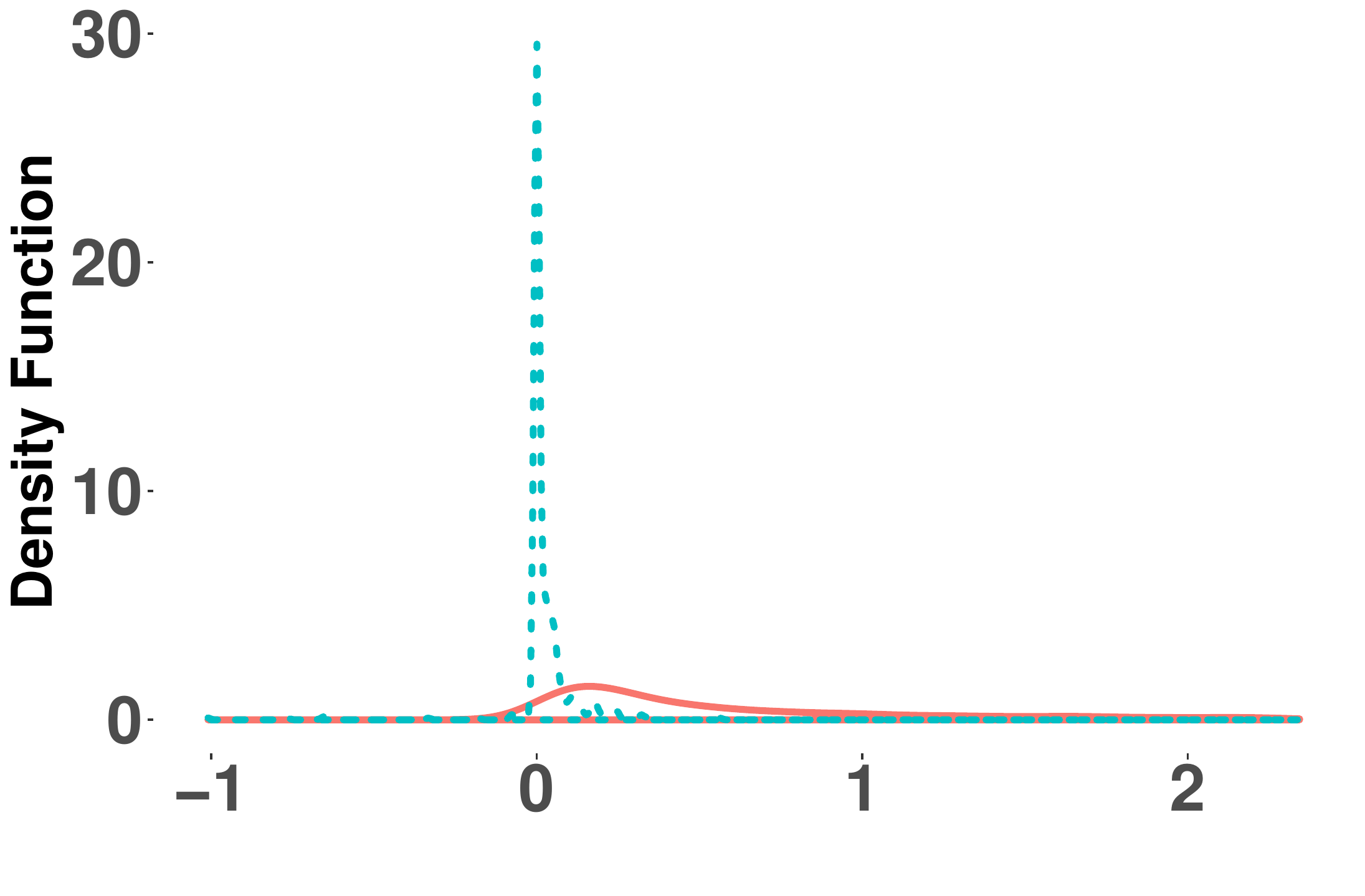}}
            \label{fig:his_div}}     
            
    \subfigure[Comment Count (CC)]
    {\fbox{\includegraphics[width=2.1in]{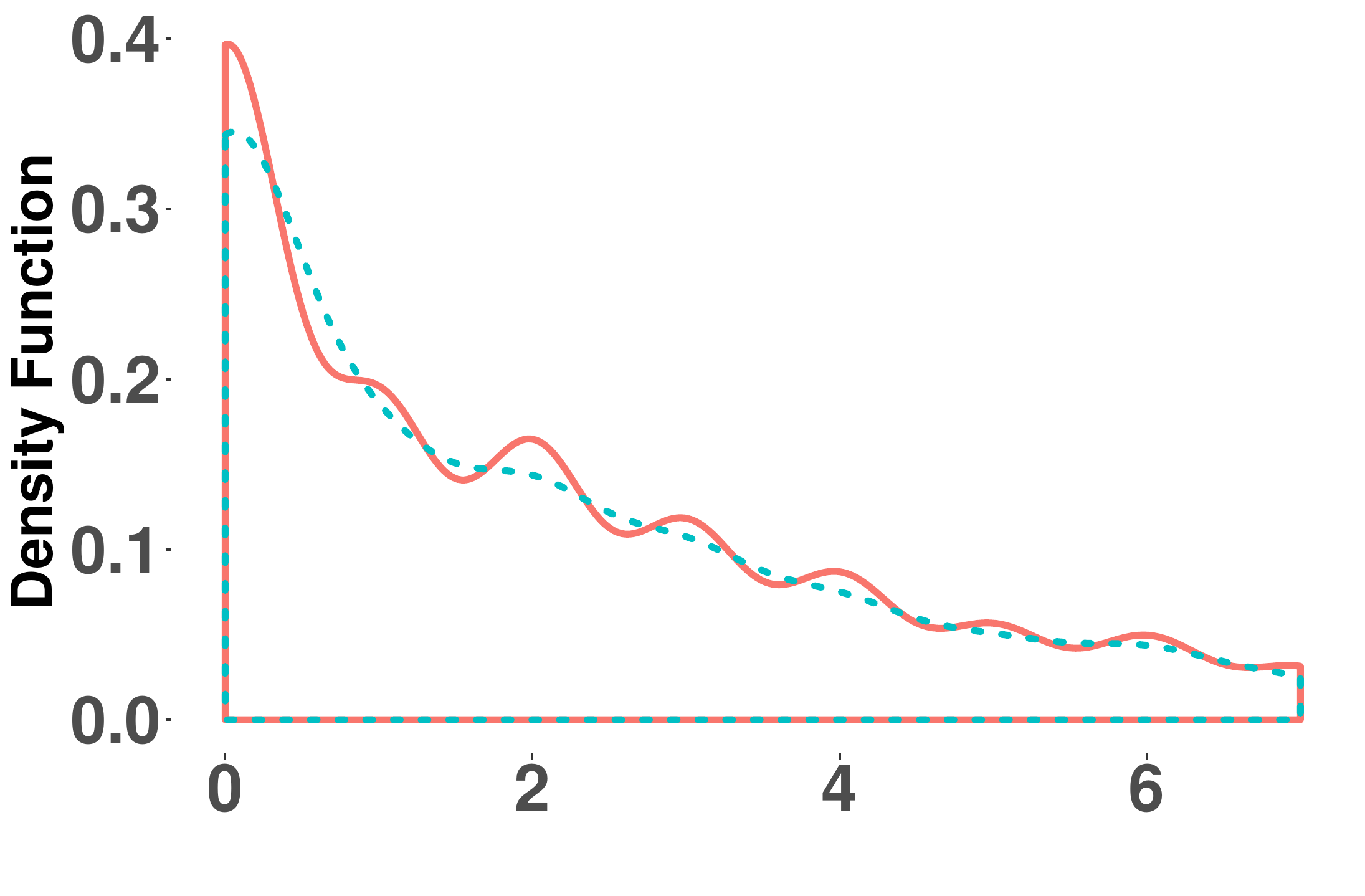}}
        \label{fig:his_comCou}}
     \subfigure[Favorite Count (FC)]
        {\fbox{\includegraphics[width=2.1in]{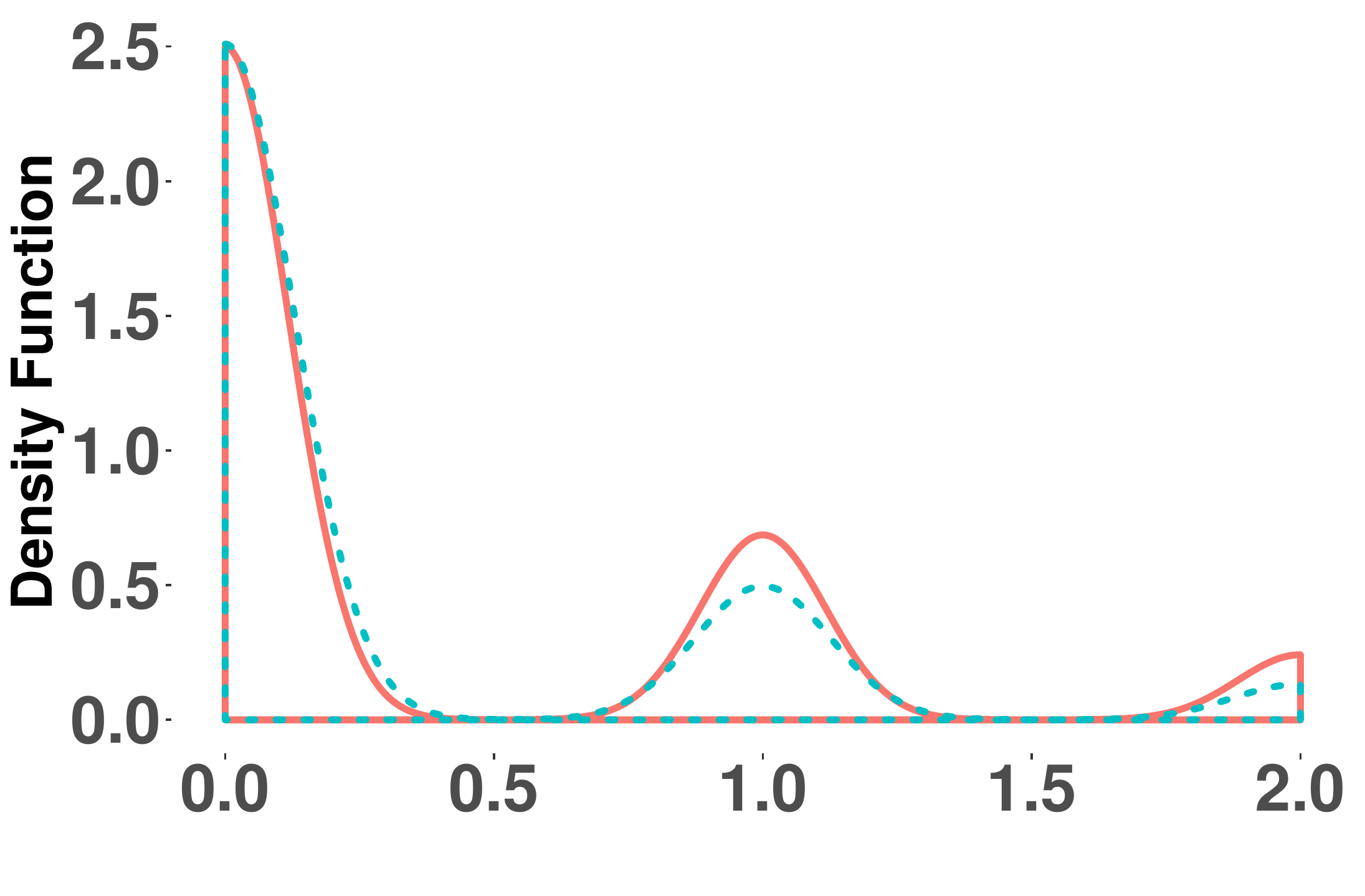}}
            \label{fig:his_favCou}}  
    \subfigure[Avg of Comment Count (ACC)]
    {\fbox{\includegraphics[width=2.1in]{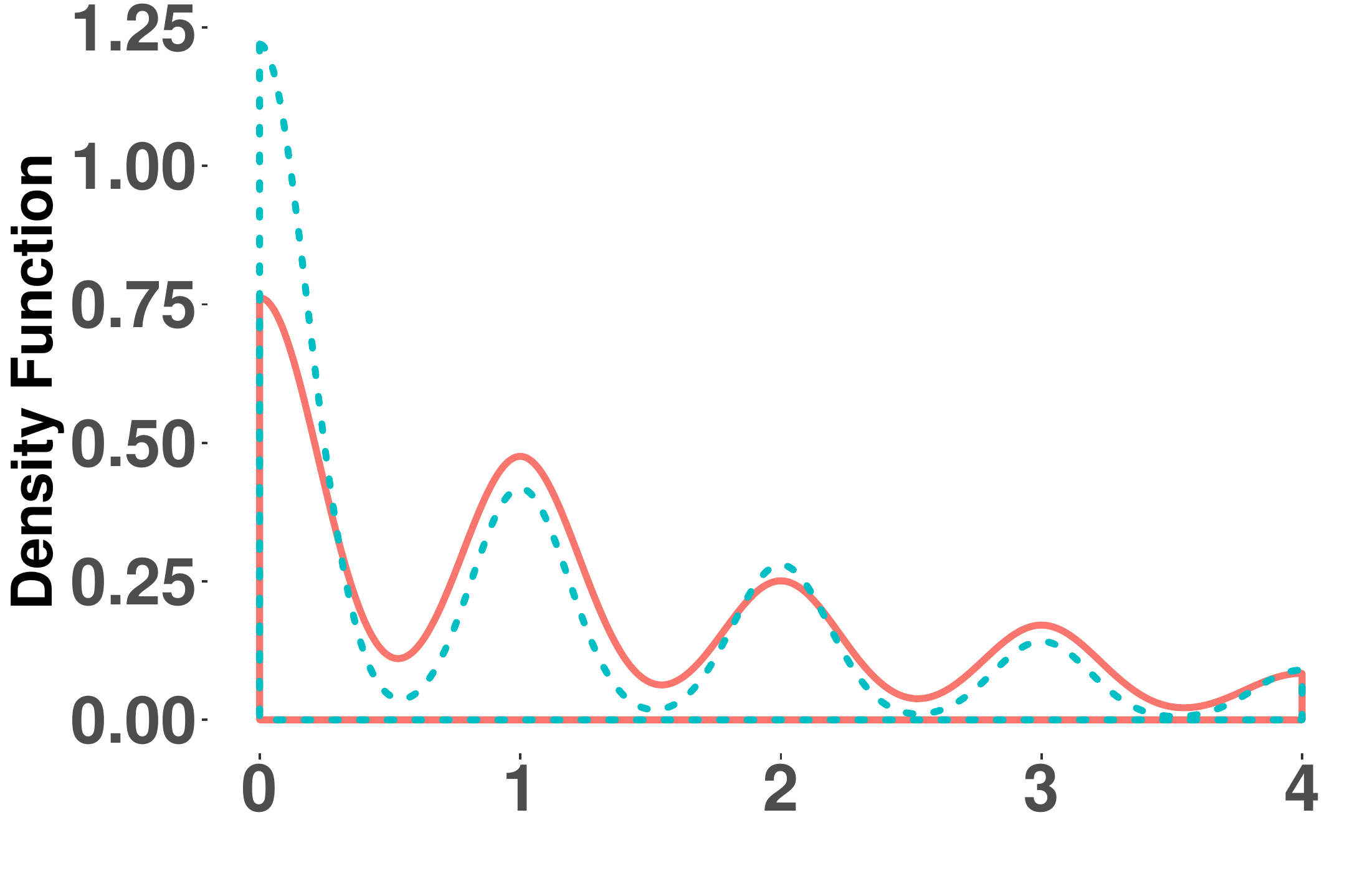}}
        \label{fig:his_avgCom}}
        
     \subfigure[Avg of Answerers' Reputation (AAR)]
        {\fbox{\includegraphics[width=2.1in]{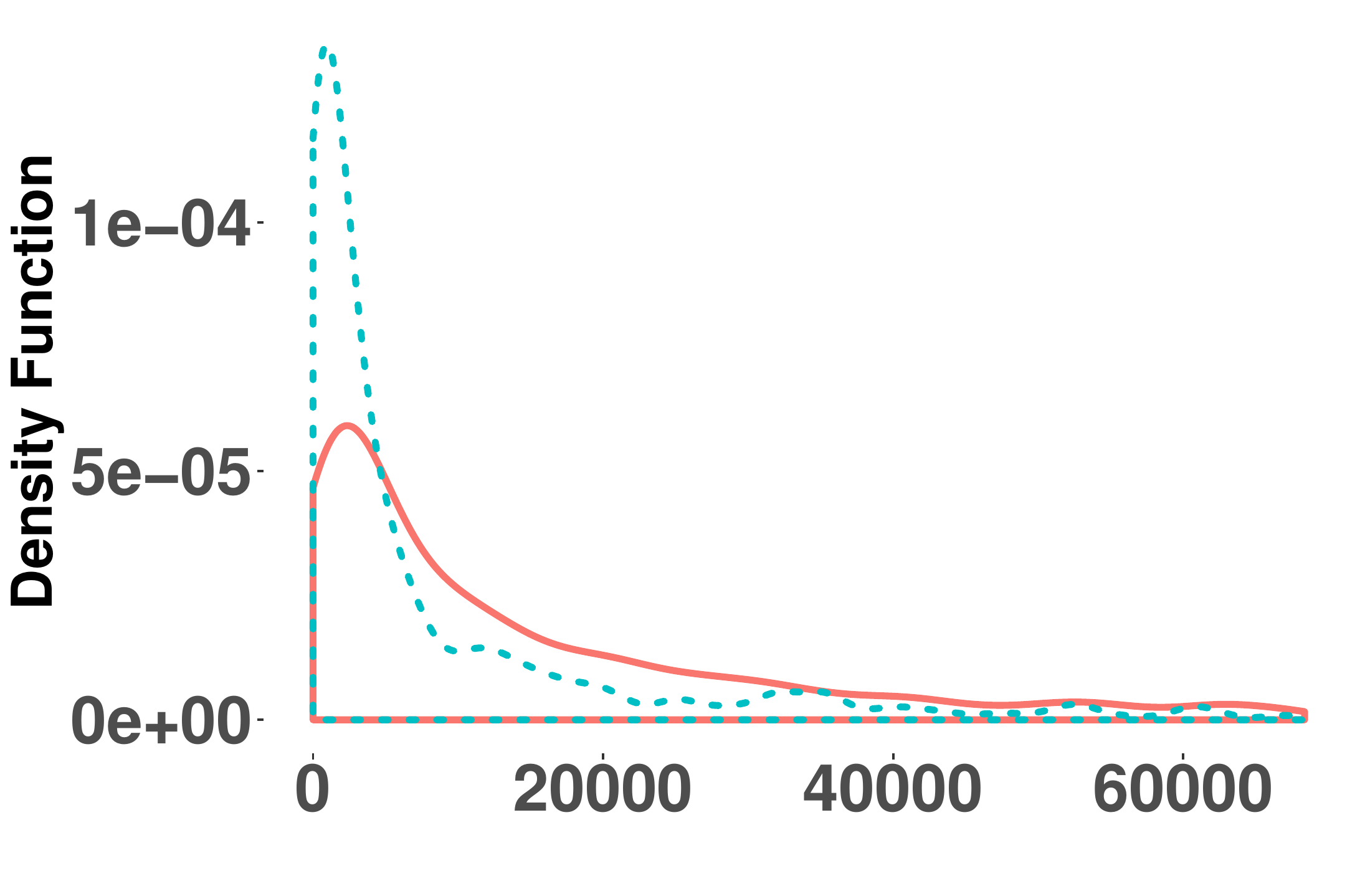}}
            \label{fig:his_ansRep}}  
      \subfigure[Asker's Reputation (AR)]
         {\fbox{\includegraphics[width=2.1in]{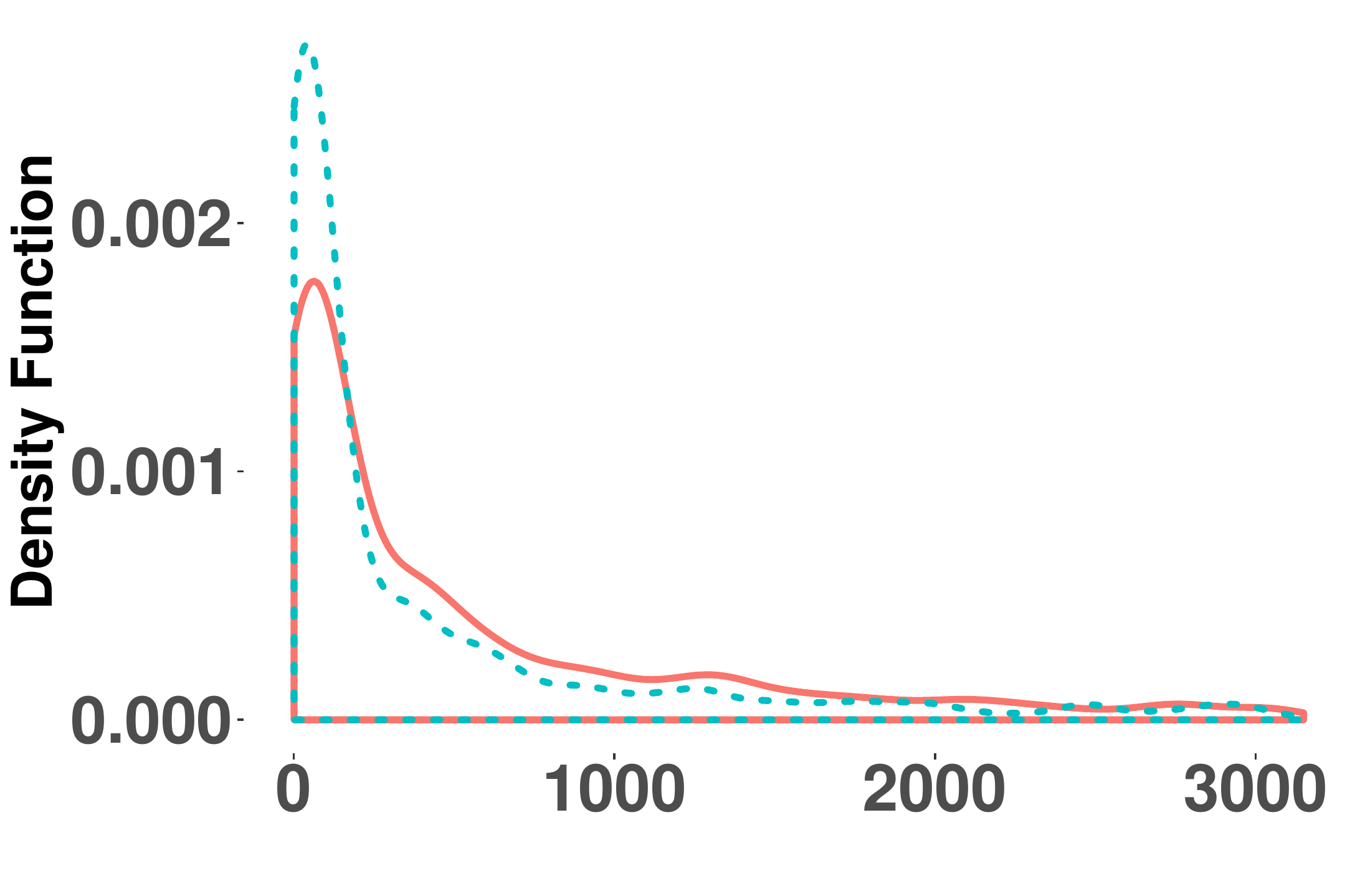}}
             \label{fig:his_askRep}}
        \setlength{\fboxsep}{2pt}
        \setlength{\fboxrule}{1pt}
        \hspace{16pt}
      \subfigure
         {\fcolorbox{teal}{pink}{\includegraphics[width=0.5in]{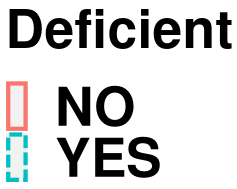}}
             \label{fig:legend}}             
    \caption{Comparing The properties of the two groups of posts, namely Deficient (YES) vs Not-Deficient (NO).}        
    \label{fig:his}
\end{figure*}

To show the difference between the two types of posts, we depicted the probability density function of the feature associated with each property of interest, and tried to find out whether they differ considerably.
The goal was to find properties that will help the model find the deficient posts more accurately.
We used Inter-quartile Ranges to remove the outliers.
Accordingly, in each feature, we removed the values below the {lower\_bound} or above the {upper\_bound}. Figure \ref{fig:his} provides multiple figures for each feature. We show deficient posts with \emph{dotted Blue} lines and the rest with \emph{simple Red} lines.
\subsection{Qualitative Analysis}
We present some insights about deficient posts in this section.
As illustrated in Figure \ref{fig:his_has}, when HAA is FALSE, the number of posts requiring improvement increases.
Figure \ref{fig:his_ansCou} shows that the probability of the deficiency would increase if the post has no answer yet. 
Although it seems that the Answer Count is an effective property, this feature does not impact the final label alone.
This is probably because answer's quality is extremely important. Also as mentioned before, there can be various solutions to a problem.
Therefore, there are posts that have more than three answers but are labeled as YES.
As shown in Figure \ref{fig:his_sco}, although the score of a question does not have a close relationship with the target class (label YES), in lower scores, the probability of needing improvement is higher than in higher scores.
Figure \ref{fig:his_sumSco} depicts the high correlation between SAS and the target class.
If we have a post with a low sum of answer scores, we may have a post that needs improvement.
As presented in Figure \ref{fig:his_viewCou}, there are several post that have been viewed many times, but need better answers. It seems this property is related to deficiency as well.
According to Figure \ref{fig:his_div}, another significant feature related to the target class is the SSVC.
If a post has a low value of this property, probably many users who view the question are not satisfied with its answers. 
If a post has a high value for this feature, it is highly probable that it has appropriate answers with many up-votes.
According to Figure \ref{fig:his_comCou}, the number of comments of a question does not have any obvious relation with the target class.
As we see in Figure \ref{fig:his_favCou}, FC of target posts are higher than the others.
However, this is not a strong relationship because FC's value is zero for many posts.
According to Figure \ref{fig:his_avgCom}, ACC cannot help our model since the figure is similar for both classes YES and NO.
As we see in Figure \ref{fig:his_ansRep}, AAR has higher values in posts that do not need improvement.  However, it does not make a noticeable difference in the figures of the two classes.
Figure \ref{fig:his_askRep} shows AR does not have any apparent relation with the target class.
\section{The Prediction Model}\label{sec:model}
In this section we will answer the following research question:

\textbf{RQ1: Can we train a model to predict deficient posts?}

To find deficient posts, we used state-of-the-art classification models with high performance in this research field.
We used feature selection methods to narrow down the set of related features to the most important ones.
A wrapper method identifies important features.
We used three wrapper methods, namely \textit{Recursive Feature Elimination},  \textit{Genetic Algorithm}, and \textit{Simulated Annealing with 10-fold cross-validation.
We concluded SAS, VC, HAA, SSVC, and AC are the most important features for our models.}
Then, we tuned each learning method with its appropriate features.
We use 80\% of the data as the training set, and the rest 20\% as the test set\footnote{Most of our implementations for creating the models using \emph{R language} can be downloaded from \url{https://github.com/MalihehIzadi/SOPI_stackoverflow_answer_quality}}.
\subsection{Decision Tree (DT)}
%
To build our predictive model, we used \textit{Leave One Out Cross Validation} (LOOC) to tune the Complexity Parameter (CP) of the DT. We tested values between infinity to 0.005 and concluded the best value for CP is equal to 0.012.
\subsection{Neural Network (NN)}
%
We utilized a neural network with one hidden layer and configured our model using LOOC. We set the number of hidden units between zero and 100.
Through our experiments, we concluded 66 units in the hidden layer leads to the best results.
\subsection{Support Vector Machine (SVM)}
%
We used the polynomial kernel and LOOC to find the best values for gamma and cost. We tested gamma values between $2^{-15}$ to $2^{-1}$ and cost values between $2^{0}$ to $2^{30}$. Eventually, we set gamma to $2^{-5}$ and cost to $2^{18}$.
\subsection{Classification Results and Evaluation}
After selecting the features and tuning each machine learning method, we ran our model on the test data.
Table \ref{tab:ml_result} provides the results of each learning method in percent.
DT's results of 94.5\% precision, 90.3\% recall, and 92.4\% F1 score indicate it performed better than the other two models, thus we used it in our prediction model\footnote{A sample image of this model is accessible at \url{https://github.com/MalihehIzadi/SOPI_stackoverflow_answer_quality}}.
To answer RQ1, we conclude our classifier can predict deficient posts with good performance.
\begin{table}[htbp]
\caption{Learning Methods' Results}
\begin{center}
\begin{tabular}{|c|c|c|c|c|c|} \hline
\makecell{Learning\\Model} & \makecell{Recall\\Sensitivity} & Precision & \makecell{Balance\\Accuracy} & Kappa & F1 \\ \hline
\makecell{DT}&\textbf{90.3\%}&\textbf{94.5\%}&\textbf{94.4\%}&\textbf{90.3\%}&
\textbf{92.4\%}\\ \hline
\makecell{NN}& 88.8\%& 92.2\%& 93.3\%& 87.9\%& 90.5\%\\ \hline
SVM& 87.3\%& 95.1\%& 93.0\%& 88.7\%& 91.0\%\\ \hline
\end{tabular}
\label{tab:ml_result}
\end{center}
\end{table}
\subsection{Feature Selection}
Decision Trees show important features when the final tree is generated.
We conclude important features are SSVC, SAS, HAA, and VC, respectively. 
The reason is that the DT tries to classify the posts using SSVC at first. Then, it uses SAS (sumScore) and HAA (accepted) for separating the posts.
According to Sec. \ref{sec:prop}, we expected the ratio of Sum of Answer Scores to the View Count (SSVC), Sum of Answer Scores (SAS) and Has Accepted Answer (HAA) to be highly related to predicting our target class.

\section{SOPI (\underline{S}tack \underline{O}verflow \underline{P}ost \underline{I}mprover)}\label{sec:appr}
In this section, we explain how SOPI works.
After users are logged in, SOPI tracks their activities in the IDE and suggests related deficient posts in need of more/better answers based on the developing context.
Users can select their desirable maximum contribution rate (e.g., once a day) in the plugin setting.
This is because we take precautions not to disturb developers or decrease their productivity while coding by excessive pop-ups or interruptions. 
Our goal is to make it simpler for the \textit{willing} ones to contribute and share their knowledge at the right time, that is when they are working on a specific context related to unanswered/poorly answered questions in SO.
It is entirely up to the developers how often (if ever) they will be prompted by the plugin.
It is needless to say in cases that the code should not be shared or analyzed in any way due to security or licensing issues, users can turn off the plugin at any moment.
Also, contributions take place in the plugin's window, not the programming windows. So they are not in the immediate viewing area of developers and hence the distraction is reduced further.

\begin{figure}
    \centering
    \fbox{\includegraphics[width = \columnwidth]{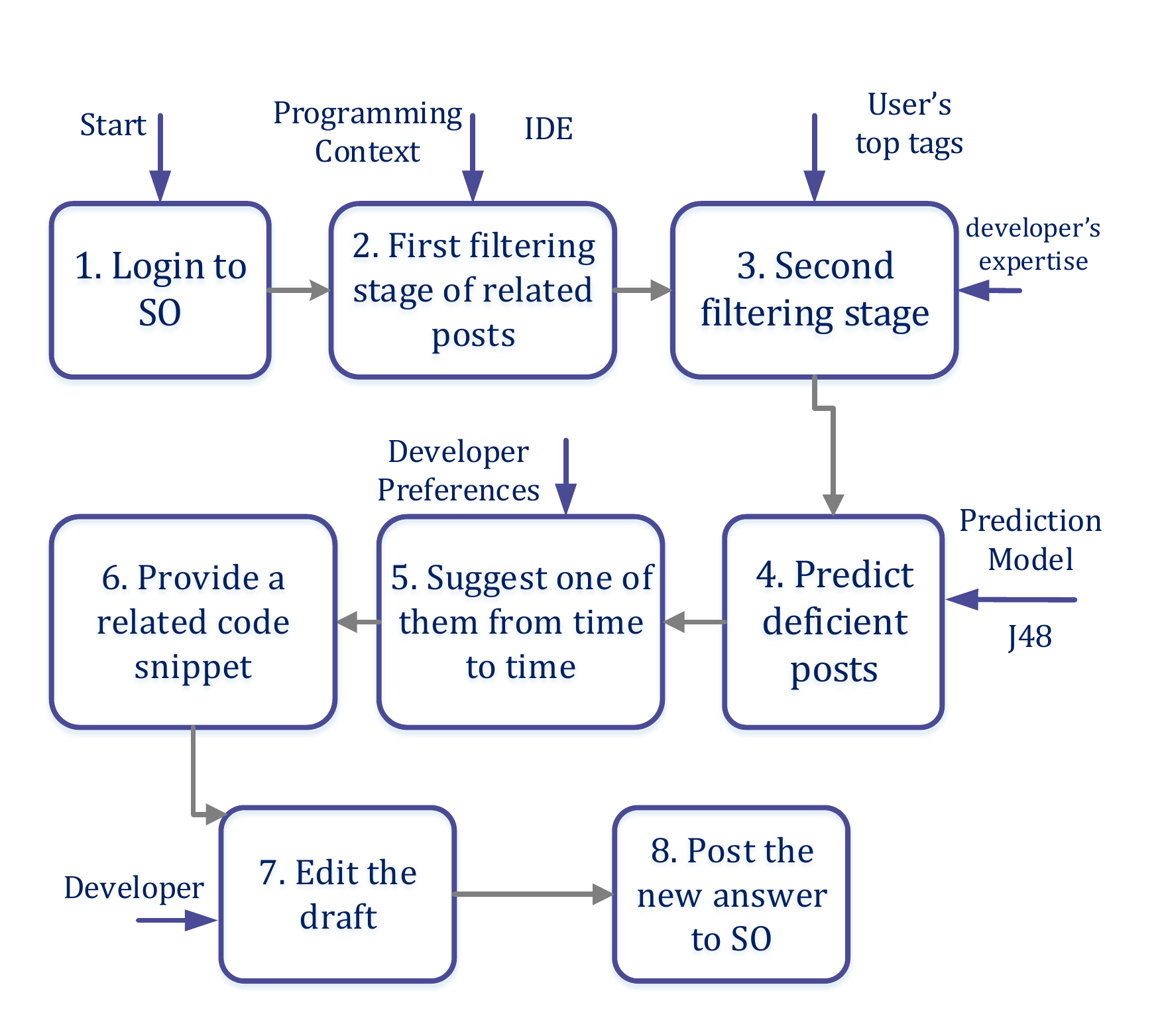}}
    \caption{The SOPI process}
    \label{fig:process}
\end{figure}

SOPI, first, interprets the programming context of the user and then filters related posts to this context. As a finer-grained filter, it selects questions from the retrieved set of posts that are related to the user's expertise as well.
SOPI determines this based on the top tags of this user in SO. Then using the DT classifier, it predicts which posts need improvement and suggests them to the developer to improve. It should be mentioned that if SOPI does not find any post after the mentioned process, it retries after a specified period which is set in the setting by the programmer.
SOPI also provides a draft answer for the developer to edit. The draft is generated using the code fragments of the question.
In the end, the improved answer is sent to SO community with minimum interrupts for the developer.
We depict this process in Figure \ref{fig:process}.
At first, we define each component of our proposed approach, and then, we integrate those components to form the whole approach.
\subsection{Identifying Related Posts}
To minimize the context switch for programmers to answer the questions, we try to find related questions according to the developer's programming context. Therefore, we use \textit{Prompter}, a tool for extracting developer's coding context \cite{ponzanelli2014prompter}.
While the authors use community-related aspects (such as reputation and scores) in conjunction with code-related aspects, in this step we focus on coding-context-related aspects such as code similarity, API similarity (types of the used API), and text similarity (e.g., names of methods invoked in the API).
This is because we intend to minimize the coding context switch as much as possible. 
This step helps refine posts based on relevant coding context.
\subsection{Identifying Programmers' Expertise}
In SO, each user has a few \emph{Top Tags} according to his previous activities. When all tags of a question exist in the set of a programmer's top tags in SO, he probably can provide good answers for those topics. Therefore, we filter the posts of the previous step based on the programmers' top tags.
This step helps refine posts based on relevant expertise as well.
\subsection{Finding Deficient Posts}
We used the DT classifier in this step as described in Sec. \ref{sec:model}.
Here we find highly-related deficient posts that the programmer can answer with high quality.
\subsection{Recommending a Code Snippet}
To facilitate the process of answering questions, we use the approach proposed by Tavakoli et al. \cite{tavakoli2016improving}.
We retrieve the code snippets of a selected question and look for similar code segments in the programmer's IDE using clone detection techniques.
To find similar code segments in the programmer's IDE we use \emph{simian-2.3.35}\footnote{\url{http://www.harukizaemon.com/simian/}} clone detection tool.
Then, we use the related code parts and create an appropriate code snippet by adding the relevant code of the programmer using Backward and Forward Slicing.
We believe this will accelerate and simplify the answering process for our cooperative developer.
To apply the slicing technique and find other relevant parts of the programmer's program, we exploit the \emph{Wala}\footnote{\url{http://wala.sourceforge.net/}} tool.
Although a good question should have code segments according to SO guidelines, if a question does not have any code segment, we will not recommend any code snippets.

Finally, we integrated all the above parts in our Eclipse plugin, SOPI.
Note that we only use posts that have not had any activities in the recent 90 days.
The reason is that the quality of a question may be improved due to the activity, but it has not affected the data about the post.
Also, we assign the posts to the programmer based on the score calculated using Prompter's approach.
That is if a selected post has 60\% similarity with the programmer's programming context, it will be assigned to her/him with a probability of 60\%.
On the other hand, with a probability of 40\%, the plugin similarly tries to assign another post to the user. This assignment process would continue until a post is assigned to the programmer.
Although our goal is to improve the posts using the help of the community and facilitating this process, we also try not to disturb the developers more than needed.
If a programmer prefers to contribute to SO daily, the plugin suggests a deficient post once a day, no more.
Finally, the plugin recommends a code snippet that the user can edit or add an explanation to, then SOPI submits the answer to SO.
\subsection{An Example of Utilizing SOPI}
Remember the post mentioned in Sec. \ref{sec:intro}.
Assume a developer with expertise in Google APIs is doing his daily programming tasks.
SOPI detects this post as a related deficient post and assigns this to the user.
Then it recommends a code snippet from the developer's IDE using clone detection and slicing techniques.
Finally, after a few edits by the developer, the code snippet shown in Figure \ref{fig:DeveloperCode} is posted to SO as a new answer to the question.
\begin{figure}
\centering
\begin{lstlisting}
final HttpTransport transport = new NetHttpTransport();
final JsonFactory jsonFactory = new JacksonFactory();
GoogleIdTokenVerifier verifier = new GoogleIdTokenVerifier.Builder(transport, jsonFactory)
   .setAudience(Arrays.asList(clientId))
   .setIssuer("https://accounts.google.com")
   .build();
GoogleIdToken idToken = null;
try {
  idToken = verifier.verify(ID_TOKEN);
} catch (GeneralSecurityException e) {
  e.printStackTrace();
} catch (IOException e) {
  e.printStackTrace();
}
GoogleIdToken.Payload payload = null;
if (idToken != null) {
     payload = idToken.getPayload();
}
String firstName = payload.get("given_name").toString();
String lastName = payload.get("family_name").toString();
\end{lstlisting}
\caption{A code snippet created by the SOPI}
\label{fig:DeveloperCode}
\end{figure}
\section{Evaluations}\label{sec:eval}
We evaluated our approach in two phases.
First, we evaluated SOPI's helpfulness for the users while contributing to SO.
Then, we evaluated the effect of our approach on the qualities of posts in SO in terms of \textit{completeness}, \textit{conciseness}, \textit{correctness} and \textit{comprehensibility} as four metrics on the quality of a set of answers.
\subsection{Evaluating the Usefulness of SOPI}
We first address the usefulness of SOPI for developers who contribute to SO while working on their code in the IDE.
The goal of this experiment is to answer the following research question:

\textbf{RQ2: ``Does SOPI facilitate the process of answering SO questions?"}

We asked 10 programmers with at least four years of Java programming industry experience to use the SOPI for three weeks, while they were doing their daily tasks.
The participants had four different role levels, namely \emph{chief technical officer} (1 participant), \emph{technical project lead} (2 participants), \emph{senior software developer} (2 participants), and \emph{software developer} (5 participants).
In the period of using SOPI, the programmers were developing a Back-end as a Service (Baas) platform such as object storage, cloud code, authentication, etc. Therefore, they faced several programming tasks such as string operations, database operations, and RESTful APIs.
This strengthened our evaluation because the participants performed various programming tasks that could cover different topics of questions on SO.
Detailed information about the participants in this experiment is available in our repository \footnote{We reported the number of years of programming experience for each participant in the last three columns. \url{https://github.com/MalihehIzadi/SOPI_stackoverflow_answer_quality}}.

We assigned deficient posts to the participants every three days and asked them to answer at least two questions.
We also designed a questionnaire and asked the participants about the similarity of assigned posts with their programming context and expertise.
Moreover, we asked them whether these posts needed improvement. If yes, according to which criteria this improvement should take place.
Finally, we asked them about the usefulness of recommended code snippet.
Table \ref{tab:eva_results} presents the results of this experiment\footnote{Detailed information can be found at \url{https://github.com/MalihehIzadi/SOPI_stackoverflow_answer_quality}}.
We report the number of average assigned posts to participants in column \emph{Assigned Count},
average number of posts related to \emph{programmer's developing context} in column \emph{Related Context} and
average number of posts related to each \emph{programmer's field of expertise} among all the assigned posts in column \emph{Related Expertise}.
The \textit{Deficient Count} column shows average number of deficient posts and the next four columns report average number of deficient posts regarding our four metrics from all participant's view.
Finally, the last two columns report average number of cases that our plugin recommended code snippets and how many of them were useful for the participant.

For instance, SOPI on average assigned 5.4 posts to each programmer from which on average 4.5 of them were related to their coding-context and so on.

\begin{table}[htbp]
\caption{Results of the SOPI Evaluation}
\begin{center}
\begin{tabular}{|c|c|c|c|c|c|c|c|c|c|c|}
\hline
\rotatebox{90}{Participant} &
\rotatebox{90}{\makecell{Assigned Count}}  &
\rotatebox{90}{Related Context} &
\rotatebox{90}{Related Expertise} &
\rotatebox{90}{\makecell{Deficient Count}}&
\rotatebox{90}{Completeness}&
\rotatebox{90}{Conciseness}&
\rotatebox{90}{Correctness}&
\rotatebox{90}{Comprehensibility}&
\rotatebox{90}{\makecell{Recommended Count}} &
\rotatebox{90}{\makecell{Useful\\Recommendations}}\\	
\hline
Avg &5.4&
\makecell{4.5}&
\makecell{5.1}&
\makecell{4.4}&
\makecell{2.4}&
\makecell{1.0}&
\makecell{0.6}&
\makecell{1.4}&
\makecell{3.8}&
\makecell{3.1}\\ \hline 
\end{tabular}
\label{tab:eva_results}
\end{center}
\end{table} 

\textbf{The Experiment's Results}:
SOPI assigned 94\% and 83\% of the posts correctly regarding relatedness to developers' expertise and programming context, respectively.
About 81\% of posts assigned by the plugin (classified by the DT model in the previous step) were determined deficient by the participants of this experiment as well.
In about 70\% of cases, SOPI suggested code snippets for the participants, from which 82\% 
facilitated the process of answering.
According to the results, it seems that utilizing code aspects used in Prompter \cite{ponzanelli2014prompter} can help assign appropriate posts to programmers.
SOPI was also successful in assigning questions to the programmers who have enough experience for answering them.
This indicates that using the history of a user's activities and question's tags are indeed effective.
However, in some cases, participants were unable to answer the assigned questions.
In further investigations, we concluded that one reason for this problem is indeed inappropriate question's tags which misled the SOPI.
SOPI was also successful in recommending code snippets to help programmers contribute to SO easier and faster.
Therefore, using the ExRec method \cite{tavakoli2016improving} can help produce good code snippets for the programmers. Note that we use the method only for questions and thus, some problems of the existing code in SO's answers, such as errors and low comprehensibility, are not included in our recommendation.
To answer RQ2, we can conclude SOPI is indeed successful in recommending relevant posts to users and at the same time provide them with good initial answers (code snippets) to facilitate answering process.

Furthermore, we analyzed how this set of participants perceived low-quality posts in terms of the four criteria we used to label deficient posts as well.
Participants believed that more than half of posts (about 54\%) need more complete set of answers.
This issue can be caused by various reasons. For instance, users tend to answer fast, maybe just to collect reputation. So they may become hasty and not address all aspects of a question being asked.
This also can make answers hard to understand, as participants claimed one-third of the posts have comprehensibility problems in their answers. Therefore, it is better that SO users take more time to comprehend the question thoroughly, and then add some details to their answers to explain their replies more clearly.
About 23\% of posts were reported to contain useless information in their answers set. This issue can occur when the answerers copy 
and paste all parts of their methods including unhelpful segments.
Participants claimed 14\% of answers were incorrect. This is probably due to the fact that answerers usually do not test their code snippets before posting them on SO and they may have syntactic or semantic errors.
Furthermore, some of the posts needed to be improved from various aspects at the same time. For example, an answer can be enhanced regarding completeness and comprehensibility areas.
Therefore we suggest that SO changes some of its policies on how it rewards users for answering questions in a way that users try to provide more comprehensive, understandable, correct and concise answers rather than just answering faster than others.

Moreover, participants were asked to determine which aspects of the post they want to improve. 
Table \ref{tab:eva_results2} reports these findings.
For instance, programmer $P_{1}$ submitted answers to three of the four deficient posts to improve completeness in two, conciseness in one, and correctness in one of the posts.
\begin{table}[htbp]
 \caption{Improvement Type based on the Metrics}
\begin{center}
\begin{tabular}{|c|c|c|c|c|c|c|}
\hline
\rotatebox{90}{Participant} & \rotatebox{90}{\makecell{Deficient\\Count}}& \rotatebox{90}{\makecell{Submit\\Count}} & \rotatebox{90}{Completeness} & \rotatebox{90}{Conciseness} & \rotatebox{90}{Correctness} & \rotatebox{90}{Comprehensibility}\\ \hline
$P_{1}$&4&3&2&1&1&0 \\ \hline
$P_{2}$&3&3&2&0&1&1 \\ \hline
$P_{3}$&6&3&2&1&0&1 \\ \hline
$P_{4}$&5&2&1&0&1&1 \\ \hline
$P_{5}$&4&2&2&0&0&1 \\ \hline
$P_{6}$&5&3&2&1&0&1 \\ \hline
$P_{7}$&4&2&1&0&0&1 \\ \hline
$P_{8}$&6&4&3&1&1&0 \\ \hline
$P_{9}$&6&3&2&1&0&1 \\ \hline
$P_{10}$&6&2&2&1&0&0 \\ \hline
\end{tabular}
\label{tab:eva_results2}
\end{center}
\end{table} 

\subsection{Evaluating the Effects of SOPI in Stack Overflow}
This experiment was designed to answer the following research question:

\textbf{RQ3: ``Do the posted answers improve the quality of questions?"}.

To find the answer to the above research question, we utilized 26 answers that had been created using our plugin in the previous experiment\footnote{The links to the answers which have been sent to SO by SOPI during our evaluation will be added for the camera-ready version.}. Next, we asked 15 other programmers with at least two years of Java programming language experience in the industry to qualify the answers.
We randomly assigned about 12 answers to each programmer in a way that each answer was validated at least seven times.
Detailed information about the participants in this experiment is available in our repository \footnote{\url{https://github.com/MalihehIzadi/SOPI_stackoverflow_answer_quality}}.

First, we asked the participants whether the posts needed improvement and concerning what metric.
Then, we asked them to investigate the effect of the new answers based on the four metrics.
Table \ref{tab:eva2_results} reports the result of this experiment\footnote{Access detailed information from our repository}.
On average, programmers claim 11.8 posts were deficient and the deficiency of 10.9 of them was resolved by utilizing the answers generated with the help of SOPI.

\textbf{The Experiment's Results}: According to our user study, about 50\% of the problems lied with the completeness of answers' set, 19\% with conciseness, 13\% with correctness, and 22\% with comprehensibility.
Therefore, to answer RQ3, SOPI improved the perceived quality of 92\% of the deficient posts on average.
\begin{table}[htbp]
 \caption{Results of the Proposed Approach Evaluation}
\begin{center}
\begin{tabular}{|c|c|c|c|c|c|c|}
\hline
\rotatebox{90}{Participant}&
\rotatebox{90}{\makecell{Deficient Count}}&
\rotatebox{90}{\makecell{Solved Count}}&
\rotatebox{90}{Completeness} &
\rotatebox{90}{Conciseness} &
\rotatebox{90}{Correctness} &
\rotatebox{90}{Comprehensibility}\\ \hline
Avg & \makecell{11.8}
&\makecell{10.9\\(92\%)}
&\makecell{5.9\\(50\%)}&\makecell{2.3\\(19\%)}
&\makecell{1.6\\(13\%)}&\makecell{2.6\\(22\%)} \\ \hline
\end{tabular}
\label{tab:eva2_results}
\end{center}
\end{table} 

\section{Threats to Validity}\label{sec:thr}
In this section, we discuss potential threats to the validity of our approach and results in four categories of Internal, External, Construct and Reproducibility threats as follows.
\subsection{Internal Validity}
The main threat to our internal validity is the set of posts collected from SO which affected the learning model. 
We tried to mitigate this threat through a random selection of a large number of posts in a prolonged time interval (2013-2019). We also covered various areas of programming in Java. 
Moreover, 85 participants were engaged in the assessment of our model. 
Logically, the characteristics of these individuals can affect the outcome of our evaluation.
To reduce this threat, we tried to get help from a large number of programmers with various skills and different experience levels. 
\subsection{External Validity}
Selecting questions of one program language (Java) to train a model can limit the expandability of our tool.
However, Java is one of the main programming languages which is widely used in practice. Furthermore, our proposed approach can be adapted for other programming languages using the variety of questions and answers in these languages available on SO or similar CQA platforms. 
Also, there are various code analyzing techniques for other languages (e.g., clone detection, and forward and backward slicing) and one can find related posts and recommend useful code snippets to programmers in a similar process.
\subsection{Construct Validity}
The main threat to our construct validity is the labeling process of posts.
We addressed this threat by assigning each post to three experts and choosing the final label based on the three generated labels. 
Also, in cases that programmers did not agree with each other, we held meetings with groups including three participants and two other programmers to reach a consensus agreement. 
The group would discuss the post and choose an appropriate label altogether.

Note that the decision of labeling a set of answers as being deficient inherently depends on people's subjective opinions.
Therefore, we intentionally used our qualitative metrics (\textit{completeness}, \textit{conciseness}, \textit{correctness}, \textit{comprehensibility}) that are subjective to some extent and can vary from one person to another. However, we clearly defined each one and communicated the definitions to the participants in face-to-face meetings to obtain more or less consistent evaluations. We also used a large number of participants with diverse backgrounds and expertise to represent a wide range of users on SO to mitigate this threat.
We also used 85 different programmers in three different levels of evaluation to mitigate these threats.
\subsection{Reproducibility}
We have shared our data set along with our source code on-line for replication by other researchers.
\section{Related Work}\label{sec:rel}
In this section, we review previous work on improvement of posts in CQA websites, specifically in SO. 
We categorize previous studies in three groups of (1) facilitating the use of SO; (2) investigating properties of SO elements such as questions, answers, and users; and (3) improving the efficiency in the processes of SO.
\subsection{Facilitating the use of SO}
Researchers have tried to help programmers use SO more easily. For example, they would develop a plugin which helps programmers browse useful and related SO posts, without the need to exit their IDEs and switch between platforms. Most of these approaches use developers' programming context and properties of posts (e.g., score, the reputation of an answerer, etc.) to recommend better pages to users \cite{ponzanelli2014prompter}.
\subsection{Investigating properties of SO elements}
Several studies try to identify important properties of SO pages, users, etc. Others search for features of unusual questions such as unanswered, closed and deleted questions with the goal to model and predict such questions \cite{asaduzzaman2013answering, correa2014chaff}. 
Some researches try to find high-quality questions, answers and code snippets posted on SO to help users find high-quality elements efficiently \cite{yang2016query,yao2018staqc,yin2018learning}. Moreover, they model such behaviors and help programmers produce high-quality posts \cite{nie2017data, calefato2018ask}. 
Finally, some studies have focused on modeling expert finding and routing questions to appropriate answerers \cite{neshati2017early}.

\subsection{Improving efficiency in the processes of SO}
The last group tries to automate the quality assessment of posts and provide solutions on how to prevent low-quality posts from being published on CQA platforms.
They use content-related properties of posts such as having code snippets, readability, and their community-related properties (e.g. reputation) to predict posts that need more review and edit before being published.
The goal is to help administrators validate the tremendous number of posts more effectively~\cite{ponzanelli2014understanding}.
Several studies also focused on investigating effects of different policies (such as badge policies) on the quality of posts \cite{chen2018data, wang2018users}.

In recent years, several approaches were introduced to help programmers find related questions and hints from SO such as \textit{Prompter}\cite{ponzanelli2014prompter}. However, none of them have tried to both facilitate the process and improve the quality of the posts which are done by SOPI.
Note that our proposed model does not necessarily compete with the previous methods.
In fact, SOPI can complement their functionality.
\section{Conclusions and Future Work}\label{sec:conc}
In recent years, SO has played an important role in solving programmers' challenges.
Although highly popular, SO has not been exploited to its highest potential by its community due to several obstacles.
The existence of a large number of posts with low-quality answers has been recognized as one of the main shortcomings of SO in this research field. These posts are either without any accepted answered or their existing answers are incomplete or incorrect. Therefore a well-organized, collective and continuous effort should be allocated to identification and improvement of these posts.
However due to time-limitations, deadline pressures and lack of incentive, programmers tend to neglect to answer these types of posts. To mitigate this issue, we introduced SOPI, an Eclipse plugin, that finds relevant posts based on developers' programming context and expertise. Then, it selects a post in SO which requires improvement. To identify such deficient posts, we trained a model using the Decision Tree approach on 3075 questions with a balanced accuracy of 94.4\%. 
Finally, SOPI recommends a code snippet to facilitate the process of generating answers for programmers and submits the new post to SO.

We evaluated our approach in two phases and confirmed its efficiency in both facilitating the process of answering questions and improving the quality of answers in SO.
It also reduces the allocated time and effort needed to do so through automating this process.

In the future, we plan to improve our model by using the content of posts. For example, the topic of a post can affect the way it is answered. That is, some topics are more specific and few users can answer them, while others are easier and more general, thus will be answered faster.
We also intend to extend the functionality of SOPI in finding highly skilled experts to assign questions to the best programmers. In particular, we will integrate more community aspects of programmers' profiles in our model as well.

\bibliographystyle{IEEEtran}
\bibliography{SOPI.bib}

\end{document}